# Generation of Laminar Vortex Rings by an Impulsive Body Force

Rabia Sonmez[1], Robert A. Handler[1,2,(a)], Ryan Kelly[3], David B. Goldstein[3], and Saikishan Suryanarayanan[4]

1) Department of Mechanical Engineering, George Mason University, Fairfax, Virginia, 22030

2) Center for Simulation and Modeling, George Mason University, Fairfax, Virginia, 22030

3) Department of Aerospace Engineering, University of Texas, Austin, TX, 78712

4) Department of Mechanical Engineering, University of Akron, Akron, OH, 44325

a) Author to whom correspondence should be addressed: rhandler@gmu.edu

It is shown that laminar vortex rings can be generated by impulsive body forces having particular spatial and temporal characteristics. The method produces vortex rings in a fluid initially at rest, and once generated, the flow field automatically satisfies the boundary conditions and is divergence-free. Numerical simulations and analytical models show that the strength of these rings can be accurately predicted by considering diffusion alone, despite the nonlinear nature of the generation process. A particularly simple model, which approximates the source of vorticity within vertical slabs, is proposed. This model predicts the ring circulation almost as accurately as a model which uses the exact geometry of the source of vorticity. It is found that when the duration of the force is less than a time scale based on the force radius and fluid viscosity, the ring circulation can be predicted accurately using an inviscid model.

# 1. Introduction

The vortex ring represents a particularly intriguing fluid dynamical flow which has attracted attention not only for its fundamental role in understanding vortex dynamics, but also for its technological applications [1]. The study of ring dynamics, however, requires that their generation be efficient, well-controlled, and repeatable. The earliest experiments [2] used the simple method of allowing a dyed water droplet to enter a water column, while modern methods [ 3-7] use well-controlled piston driven systems.

These experimental efforts have been complemented with numerical simulations which were used to study the dynamics of both laminar and turbulent vortex rings and related vortical structures  [ 8-12]. Dommermuth (1993), used a numerical method based on a Helmholtz decomposition to study vortex pair interactions with free-surfaces. The initial conditions were that of a vortex pair with a Gaussian vorticity distribution, which had to be chosen carefully to be consistent with the boundary conditions and the numerical method. Shariff et al. (1994), used finite-difference methods to study the three-dimensional instability of vortex rings based on previously developed methods [13], which also required compatibility between initial conditions and boundary conditions. Bergdorf et al. (2007) used a vortex method to simulate the evolution and decay of turbulent vortex rings using a Gaussian initial condition for the vorticity distribution, which was chosen to satisfy a divergence-free velocity field. These investigations, using both grid based and vortex methods, all require specification of initial velocity and vorticity fields consistent with boundary conditions and overall conservation of mass.

Alternate methods of generating vortex rings were used by Swearingen et al. (1995) and Smith et al. (2001), who used impulsive body forces to generate laminar vortex rings to study vortex-free surface interactions. In addition, body forces were employed to enforce no-slip boundary conditions using a feedback scheme [14]. However, in the work of Swearingen et al. (1995) and Smith et al. (2001) the details of the vortex ring generation process were not explored. For example, we would like to know the spatial and temporal structure of the impulsive force needed to produce a ring of a given strength. The goal of the present work, which focuses on the generation of laminar vortex rings, is to explore this issue using both analytical methods and direct numerical simulations using a pseudo-spectral method. The use of impulsive body forces to generate coherent vortices such as rings, has some advantages compared with methods in which initial conditions are specified. One such advantage is that by using an impulsive force, a vortex-ring may be initiated in a fluid initially at rest, and once generated, the flow field automatically satisfies the boundary conditions and is divergence-free. This may be closer, compared to other methods, to the way in which vortex rings are actually generated in experiments and therefore allows for the study of the generation process. We will show that it is possible to accurately predict the strength of the ring *a priori* based on the specific properties of body force. It is important to note that such body forces can be generated experimentally [15,16] using Lorenz forces in a conducting fluid to control wall-bounded turbulence. Although the forces generated in this way were not impulsive, there does not appear to be any reason why Lorenz forces could not be generated in this manner.

The remainder of this paper contains the following sections: *2. Governing equations and description of the force field, 3. Dimensional analysis and the evolution of the of circulation and kinetic energy, 4. Numerical Methods, 5. Numerical Simulation Results, 6. Models for the circulation of laminar vortex rings generated by an impulsive body force and comparison with numerical simulations,* and *7. Summary and Discussion*. In Sections 2 and 3 we describe the general equations of motion, a description of the temporal and spatial properties of the impulsive body force, and general equations for the evolution of the circulation and kinetic energy. The numerical approach is presented in section 4, and in section 5 we present visualizations of the flow fields, as well as results for the temporal evolution of the circulation, core radius, and kinetic energy. In section 6 we describe analytical models which can be used to predict the circulation of a vortex ring produced by an impulsive body force. Comparison of the models with numerical simulations shows that the ring circulation can be accurately predicted by considering only the diffusion of vorticity, in spite of the existence of nonlinearity in the development of the ring.

## 2. Governing equations and description of the force field

In this work we consider the motion of an incompressible Newtonian fluid governed by the momentum and continuity equations given by:

$$\frac{D\mathbf{V}}{Dt} = -\rho^{-1}\nabla p + \nu\nabla^2\mathbf{V} + \mathbf{f}, \quad (1)$$

and
$$\nabla \cdot \mathbf{V} = 0, \quad (2)$$

where $D/Dt = \partial/\partial t + \mathbf{V} \cdot \nabla\mathbf{V}$ is the material derivative, $t$ is time, $\mathbf{V} = (u, v, w) = (u_1, u_2, u_3)$ is the fluid velocity with components in the $x, y, z$, or $x_1, x_2, x_3$ directions, $p$ is the pressure, $\mathbf{f}$ is a body force per unit mass, $\rho$ is the density, and $\nu$ is the kinematic viscosity. The right-handed $x, y, z$ coordinate system will be used interchangeably with the terms streamwise, vertical or wall-normal, and spanwise respectively.

Here we are interested in a body force $\mathbf{f} = \mathbf{f}(\mathbf{x}, t) = (f_x, f_y, f_z)$ which has the following properties: (1) it is impulsive in the sense that it acts for a short time compared to other flow time scales, (2) it must generate fluid torques, and (3) It should act in a localized region of space. By *localized* we mean a region whose size is small compared to the computational domain or, in the case of a physical experiment, small compared to the experimental facility. Physical examples of body forces include those due to gravity or electromagnetic fields (Crawford and Karniadakis (1997) and Breuer et al. (2004) ) and those due to polymeric effects [17,18].

In Figure 1 we show a vertical $(y - z)$ plane of a force field which exists only in a cylindrical region located at the center of the spatial domain. This field, which has the desired properties and which we will use in this work, acts only in the $x$-direction such that $\boldsymbol{f} = \boldsymbol{f}(\boldsymbol{x}, t) = (f_x, 0,0)$. The spatial distribution of the force is described by:

$$f_x = A(t) g_r(r) g_x(x), \qquad (3)$$

where

$$g_r = \frac{1}{2}(1 + \tanh(\beta[R - r])), \qquad (4)$$

and

$$g_x = \frac{1}{2}(1 + \tanh(\beta[l_f/2 - |x - x_0|])). \qquad (5)$$

In these expressions, the hyperbolic tangent function is used to smooth the force field at its edges by using the tapering parameter $\beta$. In these expressions, $R$ and $l_f/2$ define the nominal radius and length of the cylindrical region, $r = \sqrt{(y - y_0)^2 + (z - z_0)^2}$, and $x_0, y_0,$ and $z_0$ define the location of the center of the force field. In all cases described here, $x_0$, $y_0$, and $z_0$ denote the center of the domain. The force is turned on or off using the function $A(t)$ which is defined by:

$$A(t) = \begin{cases} F_0, & 0 \leq t \leq \tau \\ 0, & t > \tau \end{cases}, \qquad (6)$$

where $F_0$ is the time independent force amplitude, and $\tau$ is the temporal duration of the force. In Figure 2, the geometry of the force field is shown schematically in a horizontal $(x - z)$ plane. It will be shown that the force field defined here can generate vortex rings that propagate in the $x$-direction.

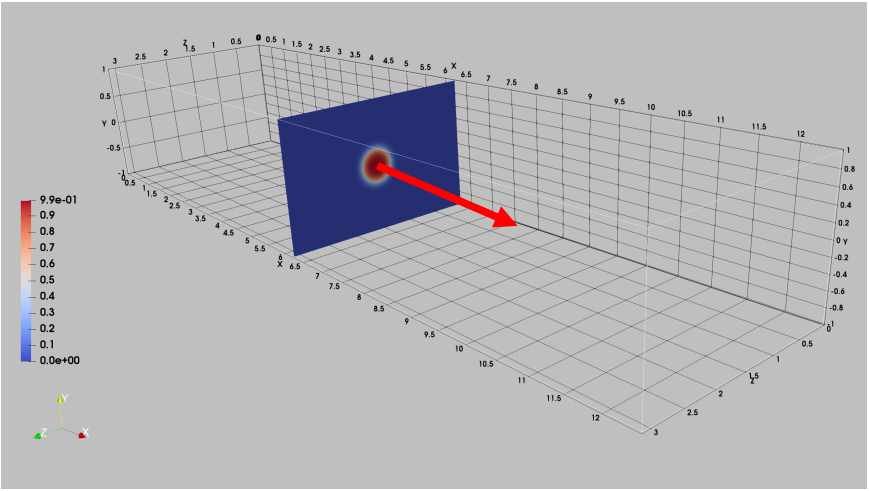

**Figure 1**: A cross-section in the $y-z$ plane of the $x$-component of the body force used to generate vortex rings is shown. The force exists in a cylindrical region as described by equations (3-6). The color-bar corresponds to the $x$-component of the force, $f_x$, made non-dimensional by $F_0$ (see equation 6). The arrow indicates that the force acts in the positive $x$ direction. The coordinates $x, y,$ and $z$ are given here in centimeters. The domain lengths in the $x$ and $z$ directions are $l_x = 4\pi$ cm and $l_z = \pi$ cm, and in the $y$-direction, the half-height is $l_y = 1$ cm. The domain size made non-dimensional by $l_y$ is $L_x = 4\pi$, $L_y = 2$ and $L_z = \pi$. The force is computed using $128 \times 65 \times 128$ grid nodes in the $x, y,$ and $z$ directions respectively. In all simulations reported here, the length and radius of the cylindrical region defining the force were $l_f = 2R = 0.5$ cm, and the force was applied such that its center corresponded to the center $(x_0, y_0, z_0)$ of the computational domain. In addition, the tapering parameter used to define the body force was set to $\beta = 15 cm^{-1}$, which gives a reasonably sharp edge to the force field. The entire computational domain is shown in the image.

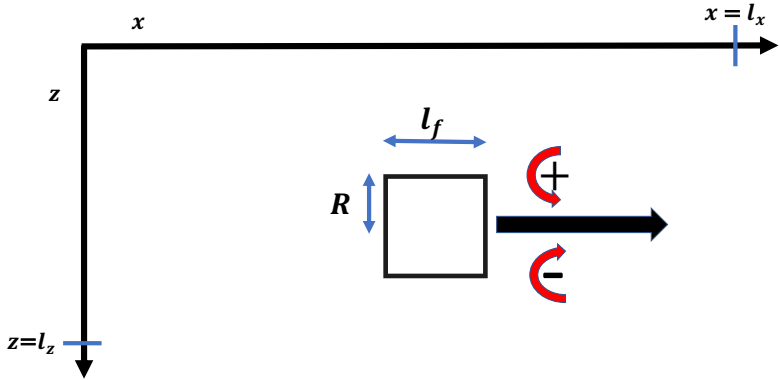

**Figure 2**: Schematic (not drawn to scale) showing the region in the $x-z$ plane in which the cylindrical body force, $f_x$, acts. The center $(y = 0)$ plane is shown. The length and radius of the cylindrical region are given respectively by $l_f$ and $R$. The center of the box is located at the exact center $(x_0, y_0,$ and $z_0)$ of the computational domain. The rightward directed arrow indicates the direction in which the force acts, and red arrows indicate the sense of rotation of the vortex ring. The $y$-coordinate is directed outward and perpendicular to the plane of the paper.

# 3. Dimensional analysis and the evolution of the circulation and kinetic energy

### 3.1 Dimensional analysis

An overall description of the problem considered here can be gained by performing a dimensional analysis. For this purpose, it is useful to divide the problem into two time periods, the generation period (T1) for which $0 \leq t \leq \tau$, and the decay period (T2), for which $t > \tau$. In T1 the body force acts on the fluid generating the vortex ring, and in T2 the kinetic energy of the ring is dissipated. In the following analysis we will use the circulation, $\Gamma$, of the vortex ring as the dependent variable, where $\Gamma = \oint \mathbf{V} \cdot d\mathbf{l}$, is the line integral of the velocity taken around an appropriately defined closed curve. The choice of the ring circulation as the dependent parameter follows from the work of Glezer (1981,1988) who found that $\Gamma/\nu$ is important in characterizing the vortex ring strength, and determining, for example, whether the ring remains laminar. It follows that an appropriate description for the circulation of the vortex ring in T1 is given by:

$$\Gamma = A(F_0, l_f, \beta, R, \nu, t) \quad . \qquad (7)$$

If we choose a velocity scale $v^* = \sqrt{F_0 l_f}$ and a time scale $\tau^* = \frac{l_f^2}{\nu}$ then equation (7) can be written in the dimensionless form:

$$\Gamma/\nu = B\left(Re_f, t^*, \frac{R}{l_f}, \beta R\right) , \qquad (8)$$

where $Re_f = \frac{v^* l_f}{\nu}$ is a Reynolds number based on the body force amplitude, and $t^* = \frac{t\nu}{l_f^2}$ is a dimensionless time. We also note that $\Gamma/\nu$ is sometimes referred to as the vortex ring Reynolds number, $Re_\Gamma$ [11]. In T2 the temporal duration of the body force, $\tau$, must be included in the analysis, leading to:

$$\Gamma/\nu = B\left(Re_f, t^*, t_{Ro}, \frac{R}{l_f}, \beta R\right), \qquad (9)$$

where $t_{Ro} = \frac{\tau \nu}{l_f^2}$, which is referred to as the Roshko number [19].

### 3.2 Generation of vorticity and circulation

Some distinction should be made concerning the physical process by which a body force generates vorticity, and the way vorticity is typically generated experimentally. As previously mentioned, vortex rings are generated experimentally by using a piston to drive fluid through a cylindrical tube. The fluid motion within the tube generates vorticity at the inner surface of the cylinder, which is incorporated within the vortex ring which exits the tube. In contrast, when a spatially compact body force acts on the fluid, the vorticity which is generated diffuses immediately, at the time of its creation, into the body of the fluid. This difference should be kept in mind in the descriptions which follow.

The manner in which a body force can be used to generated vorticity can be understood by examining the equation for the evolution of the vorticity $\boldsymbol{\Omega} = \boldsymbol{\nabla} \times \boldsymbol{V} = (\Omega_x, \Omega_y, \Omega_z)$ for an incompressible fluid of constant density given by:

$$\frac{D\boldsymbol{\Omega}}{Dt} = (\boldsymbol{\Omega} \cdot \boldsymbol{\nabla})\boldsymbol{V} + \nu \nabla^2 \boldsymbol{\Omega} + \boldsymbol{\nabla} \times \boldsymbol{f}, \qquad (10)$$

where the term on the left-side is the material derivative of the vorticity, and on the right-side the first term represents stretching and tilting of vorticity by velocity gradients, the second term represents diffusion of vorticity by viscosity, and the third term represents the generation of vorticity by the body force per unit mass, $\boldsymbol{f}$. Equation (7), together with the definition of the body force equations (3-6), can be used to determine expressions for the time rate of change of the vorticity at time $t = 0$, assuming the fluid is initially at rest (see Appendix A). This results in:

$$\left.\frac{\partial \Omega_y}{\partial t}\right|_{t=0} = \frac{\partial f_x}{\partial z}, \quad and \quad \left.\frac{\partial \Omega_z}{\partial t}\right|_{t=0} = -\frac{\partial f_x}{\partial y}. \qquad (11)$$

This shows that the force field generates both spanwise ($\Omega_z$) and vertical ($\Omega_y$) vorticity. To determine the circulation around the ring, it will be convenient to integrate the vertical component of vorticity over a horizontal ($x - z$) plane. According to equation (8), the source of this component of vorticity is $\frac{\partial f_x}{\partial z}$, which is visualized in Figure 3. We also note that since vortex lines cannot end in a fluid, the generation of both vertical and spanwise components of vorticity must form circular vortex lines characterizing a vortex ring.

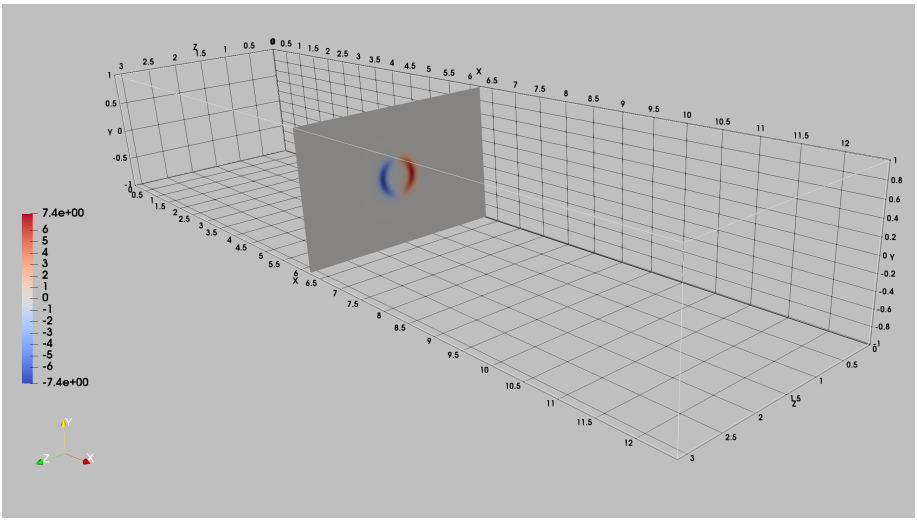

**Figure 3**: A cross-section in the $y - z$ plane of source of the vertical ($y$) vorticity is shown. The color-bar corresponds to $\frac{\partial f_x}{\partial z}$ made non-dimensional by $\frac{F_0}{l_y}$. The source is seen to be composed of positive and negative crescent shaped regions which derive from the cylindrical geometry of the force field.

In order to define the circulation for our problem, we note that we are primarily interested in laminar vortex rings which have symmetry with respect to the plane $z = l_z/2$, as well as symmetry with respect to the center plane, $y = 0$, where $l_x$, $2l_y$, and $l_z$ are the domain lengths in the $x$, $y$, and $z$ directions respectively as shown in Figure 1. These symmetries are imposed on the flow by the force field. We also note that such vortex rings must retain these symmetries indefinitely unless they become unstable. Such considerations lead to the following definition for the circulation, which follows from Stokes theorem:

$$\Gamma = \overline{\Omega_y(x,0,z,t)} \equiv \int_0^{l_z/2} \int_0^{l_x} \Omega_y(x,0,z,t) dx dz , \qquad (12)$$

where $\Omega_y(x, 0, z)$ is the vertical component of vorticity on the center plane, and an overbar is defined as the integral over the half-plane given by the region $0 \leq z \leq l_z/2$ and $0 \leq x \leq l_x$. An equation for the evolution for the circulation can be derived by first taking the curl of the momentum equation written in rotation form, which results in the following equation for the vertical component of vorticity:

$$\frac{\partial \Omega_y}{\partial t} = \frac{\partial H_x}{\partial z} - \frac{\partial H_z}{\partial x} + \nu \nabla^2 \Omega_y + \frac{\partial f_x}{\partial z} , \qquad (13)$$

where $H_x = v\Omega_z - w\Omega_y$, and $H_z = u\Omega_y - v\Omega_x$. We next apply the overbar operation to equation (10) using all flow symmetries and taking into account the periodic boundary conditions in the $(x - z)$ plane, which are used in all simulations in this work. This results in an equation for the evolution of the circulation (see Appendix B):

$$\frac{\partial \Gamma}{\partial t} = D + L + \overline{\frac{\partial f_x(x,0,z)}{\partial z}} , \qquad (14)$$

where $D = \nu(\overline{\Omega_y})_{yy}\big|_{y=0}$ represents the viscous diffusion of vertical component of vorticity and $L = \nu \int_0^{l_x} [\frac{\partial \Omega_y}{\partial z}\big|_{z=L_z/2} - \frac{\partial \Omega_y}{\partial z}\big|_{z=0}]\big|_{y=0} dx$ represents the net lateral flux of vertical vorticity.

It is important to note that when the fluid viscosity is zero and the tapering parameter $\beta \to \infty$, equation (14) can be integrated in space according to equation (12) and in time in the interval ($0 \leq t \leq \tau$) to give:

$$\Gamma = F_0 l_f \tau . \qquad (15)$$

The expression given by equation (15) represents the maximum possible circulation which we call the *inviscid circulation limit*. In this limit, the circulation will increase linearly in time while the force is active. For a fluid with viscosity, since $D$ and $L$ are each expected to be negative, the circulation should increase more slowly than linearly in time according to equation (14). It follows that the vortex ring Reynolds number can never exceed $Re_\Gamma = F_0 l_f \tau / \nu$.

### 3.3 Evolution of the volume-average kinetic energy

In addition to the circulation, the volume-average kinetic energy per unit mass given by $K = \frac{1}{V}\int k\,dv \equiv <k>$, where $k = \frac{u_i u_i}{2}$ where the integral is taken over the entire flow volume, $V$, is a particularly useful measure of the flow dynamics since its evolution equation has a simple physical interpretation. We will therefore use it to give an overall impression of the flow. The evolution equation for $k$ follows from equation (1) (See Appendix E):

$$\frac{\partial k}{\partial t} = -\partial_j(k u_j) + \rho^{-1}\partial_j(p u_j) + \nu \partial_j \partial_j k - \nu \partial_j u_i \partial_j u_i + f_j u_j \quad . \quad (16)$$

Performing volume averaging on equation (7) and using all boundary conditions, gives the evolution equation for $K$ as follows:

$$\frac{\partial K}{\partial t} = -\epsilon + <f_j u_j> . \quad (17)$$

where the dissipation of kinetic energy is given by $\epsilon = \nu <\partial_j u_i \partial_j u_i>$ and $<f_j u_j>$ can be identified as the work done on the flow by the applied body force. According to equation (17) at the instant the body force is turned off the kinetic energy must decay ($\frac{\partial K}{\partial t} < 0$) since the dissipation is strictly positive.

## 4. Numerical Methods

To determine the extent to which impulsive body forces of the kind described above can generate vortex rings with the desired characteristics, a series of direct numerical simulations were performed. For this purpose, equations (1,2) were solved using a pseudo-spectral method [17,18] in a channel geometry in which all field variables are expanded in Fourier modes in the horizontal $(x - z)$ plane and in Chebyshev modes in the vertical direction. Spectral methods exhibit exponential convergence [20] and have been used with success in simulating fully developed turbulent flows [21,22]  The computational domain lengths were $L_x = 4\pi$, $L_y = 2$ and $L_z = \pi$ in the $x, y,$ and $z$ directions respectively, where these lengths have been made non-dimensional by the channel half-height, $l_y$. In these simulations, vortex rings will travel in the positive $x-$ direction since the body force always acts in the positive direction as shown in Figure 1. There were $128 \times 65 \times 128$ grid nodes in the $x, y,$ and $z$ directions respectively. In terms of physical dimensions, the domain lengths in the $x$ and $z$ directions were $l_x = 4\pi$ cm and $l_z = \pi$ cm, and in the $y$-direction, the half-height was $l_y = 1$ cm. The corresponding distances between grid nodes were $\Delta x = 9.82\times 10^{-4}$m and $\Delta z = 2.45 \times 10^{-4}$m respectively. In the $y$-direction, the use of Chebyshev polynomials results in a nonuniform grid. The smallest grid resolution occurs at the boundaries of the computational domain where $\Delta y = 1.2 \times 10^{-5}$m, and at the center of the domain the resolution was $\Delta y = 4.91 \times 10^{-4}$m. The grid resolution is therefore sub-millimeter in all three directions. In all simulations the fluid was initially at rest, $V(x, t = 0) = 0$. On the top and bottom walls of the channel, shear-free conditions, $\partial u_1 / \partial x_2 = \partial u_3 / \partial x_2 = u_2 = 0$, were imposed on the velocity field, as well as $\partial \Omega_2 / \partial x_2 = 0$ on the vertical vorticity. In all simulations reported here, the length and radius of the cylindrical region defining the force were $l_f = 2R = 0.5$ cm, and the force was applied such that its center corresponded to the center $(x_0, y_0, z_0)$ of the computational domain. In addition, the tapering parameter used to define the body force was set to $\beta = 15\ cm^{-1}$, which gives a reasonably sharp edge to the force field (see Figures 1-3). In addition, the time step was $\Delta t = 10^{-4}\ s$, the body force acted for $\tau = 0.2\ s$, and the duration of each simulation was 2.0 s. All relevant symbols used in this paper are listed in Table1 in Appendix C.

A series of 12 simulations were performed in which the force amplitude $F_0$ and the kinematic viscosity, $\nu$, were varied. Three kinematic viscosities were chosen ($\nu = 10^{-2}, 10^{-1}, 10^0$ in units of $cm^2/s$) so as to achieve force-based Reynolds number, $Re_f$, which span two orders of magnitude from 3.535 to 353.5. With this choice of parameters, for each choice of viscosity, the maximum inviscid ring Reynolds number based on circulation was $Re_\Gamma = 10^3$. The parameters for all runs are given in Table 2 in Appendix D.

## **5. Numerical Simulation Results**

One of the main objectives of this work is to develop models which can predict the circulation around laminar vortex rings generated by an impulsive body force. The circulation can be used to determine the ring Reynolds number, $Re_\Gamma$, which in turn can be used to predict whether the ring may ultimately become a turbulent. Although there is no clearly defined Reynolds number at which rings become turbulent, the data compiled by Glezer (1988) indicates that turbulent rings form when $Re_\Gamma \gtrsim 10^4$ or slightly below this. To motivate the development of such models, it is useful first to examine the results of the direct simulations. To this end, we present two and three-dimensional flow visualizations, the temporal evolution of the circulation and the flow kinetic energy, and the evolution of the equivalent ring core radius.

### **5.1 Two and three-dimensional flow visualizations**

In Figure 4 the evolution of the vertical vorticity in the $x-z$ plane at the center of the domain ($y=0$) is shown for the case $\nu = 10^{-2}\ cm^2-s^{-1}$, and $F_0 = 99.95\ cm-s^{-2}$. The images are shown at four instants. Figures 4(A) and 4(B) are obtained from T1 in which the body force acts on the fluid and Figures 4(C) and 4(D) are from T2 in which the force does not act. Figure 4(A) shows the vertical vorticity field after one time step, where it should be recalled that the fluid was initially at rest. Here it is evident that the vorticity is composed of two *sheets* which appear as positive (red) and negative (blue) lines. The sheets (in the $x$-direction) are seen to have lengths $l_f$, the linear dimension of the cylindrical force field shown in Figure 1 and each sheet is a distance $R/2$ from the centerline at $z = L_z/2$. These vortex sheets are a natural consequence of the shear flow generated by the action of the force field. In Figure 4(B) the vortex sheet is seen to thicken laterally and the vorticity becomes more concentrated with a shape more reminiscent of a ring. In Figure 4(C), the force has ceased to act on the fluid, and the vorticity becomes further concentrated into positive and negative *cores*. Finally, in Figure 4(D), although the cores are still evident, the vortex strength is seen to have decreased relative to Figure 4(C), and the vorticity field is obviously diffusing laterally. A three-dimensional sequence of images of the same case shown in Figure 4 is given in Figure 5. In each image an iso-surface of the vorticity magnitude is shown together with selected velocity vectors. Since both $y$ and $z$ components of vorticity are present, ring shaped iso-surfaces are formed.

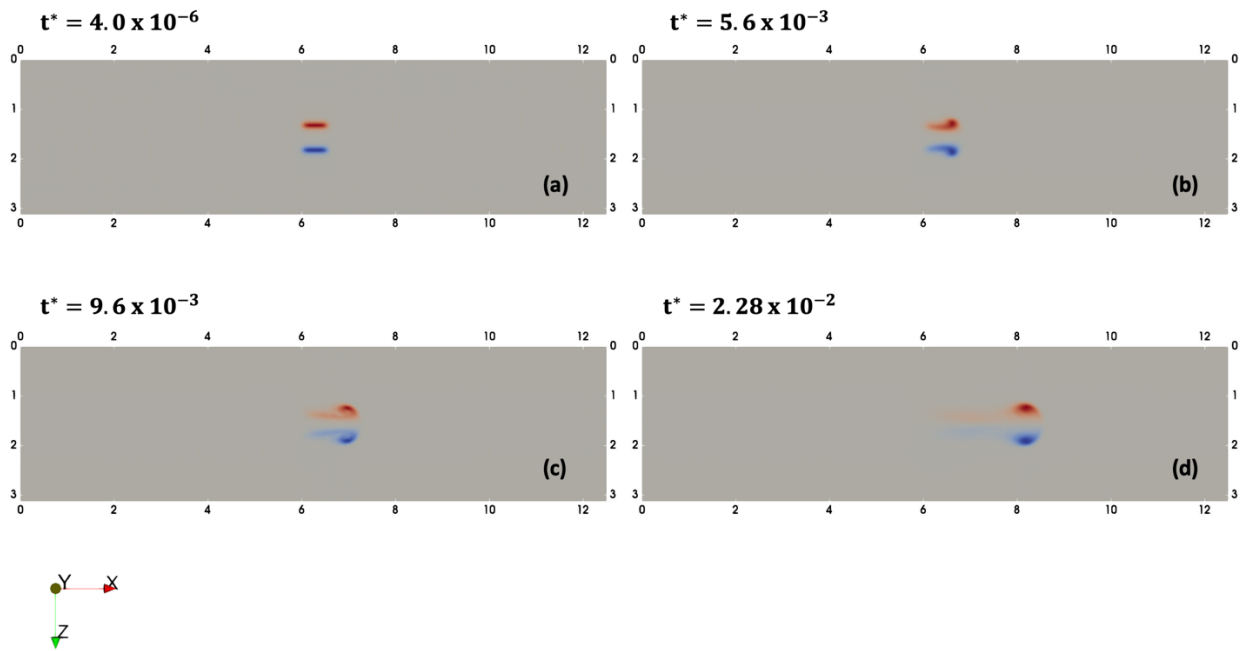

**Figure 4**: Sequence of two-dimensional snapshots of the evolution of the vertical vorticity (red=positive, blue=negative) for the case in which $\nu = 10^{-2}\ cm^2/s$, and $F_0 = 99.95\ cm/s^2$. Maximum vertical vorticity in $s^{-1}$; (A) $7.4 \times 10^{-2}$, (B) $8.4 \times 10^1$, (C) $9.7 \times 10^1$ and (D) $6.1 \times 10^1$. It should be noted that T1 ends at approximately $t^* = 8.0 \times 10^{-3}$. The visualizations are shown in an $x - z$ plane at the center of the domain ($y = 0$). The entire horizontal plane is shown in the images. A movie of the full sequence can be found at https://youtu.be/JOJy44cHrvk.

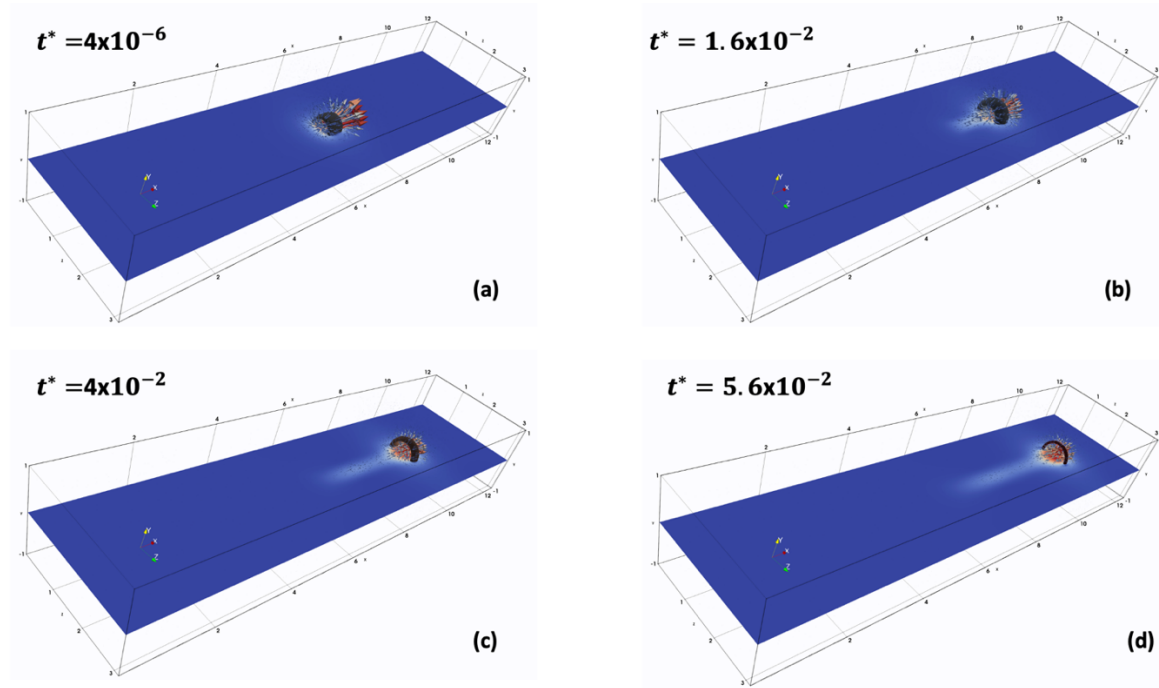

**Figure 5**: Sequence of three-dimensional snapshots of the vorticity magnitude and velocity vector fields for the case shown in Figure 4. Iso-surface of vorticity magnitude in $s^{-1}$; (A) $3.74 \times 10^{-2}$, (B) 30, (C) 30 and (D) 30. A movie of the full sequence can be found at https://youtu.be/lfxnsm6rVml.

## 5.2 Evolution of the circulation, equivalent core radius, and the kinetic energy

In Figure 6 the entire temporal evolution of ring circulation (regions T1 and T2) corresponding to the visualizations of Figures (4) and (5) are shown. Each curve corresponds to a different force amplitude. It is evident that during the time the force acts that the circulation increases approximately linearly in time and subsequently decays after removal of the force. Furthermore, it appears that the maximum circulation increases approximately linearly with $F_0$. These results appear to be consistent with the inviscid result, $\Gamma = F_0 l_f t$, indicating that the viscosity is sufficiently low to allow such an approximation. Following equation (8), the inviscid result can be put in non-dimensional form as follows:

$$\Gamma/\nu = Re_f^2 t^* \quad , \tag{18}$$

where equation (18) is assumed to be valid for short time scales in which the effects of viscosity are negligible.

This suggests that in general $\Gamma/\nu = Re_f^2 T(t^*)$ should be valid for longer time scales and higher fluid viscosities and furthermore that $\Gamma^* = T(t^*)$ where $\Gamma^* = Re_\Gamma Re_f^{-2}$. To determine the validity of this scaling, we show in Figure (7) the results of $\Gamma^*$ versus $t^*$ for all simulations. These results show that, to a large extent, $\Gamma^* \sim T(t^*)$, although some variation from this scaling at higher values of $t^*$ is evident. In this context, it is important to note that this scaling of the circulation and time gives excellent data collapse despite a two order of magnitude variation in $Re_f$. It is also clear from Figure (7) that $\Gamma^*$ does not vary according to equation (18) for the large values of $t^*$ which indicates that a complete model for the evolution of the vortex ring circulation must include the effects of viscosity.

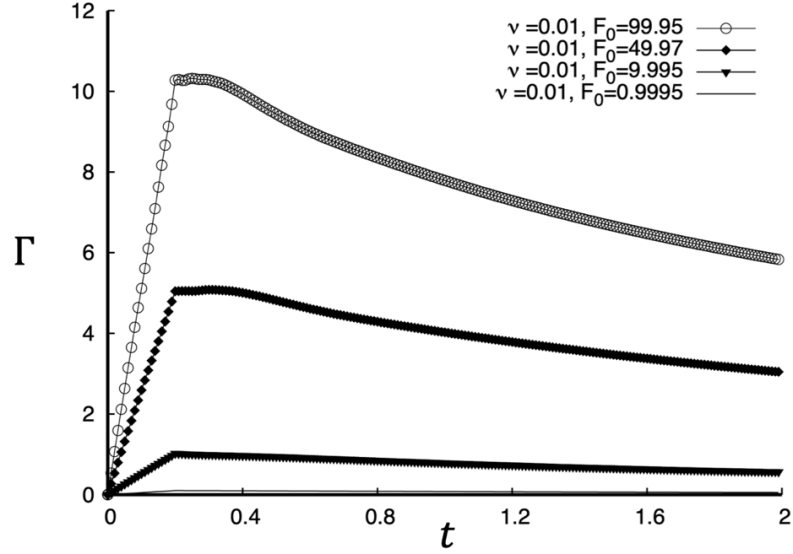

**Figure 6**: Circulation, Γ, in $cm^2/s$ versus time (s) for the case in which $\nu = 10^{-2}\ cm^2/s$, and for various $F_0$ values shown in the graph. The force acts in the time interval $0 \leq t \leq 0.2s$.

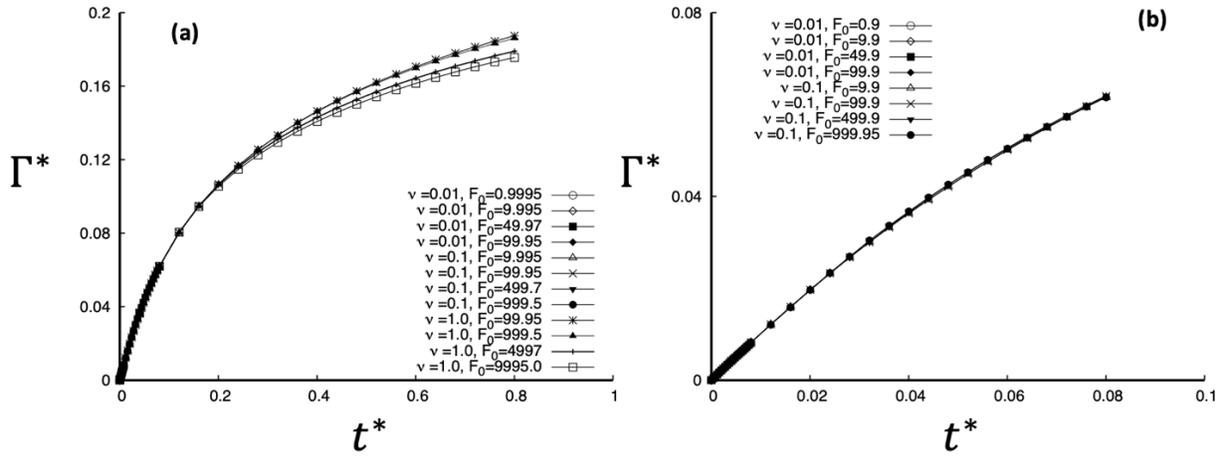

**Figure 7**: Non-dimensional circulation, $\Gamma^* = Re_\Gamma Re_f^{-2}$, versus non-dimensional time, $t^* = \frac{t\nu}{l_f^2}$ in T1. (A) all cases, (B) cases $\nu = 10^{-2}\ cm^2/s$ and $\nu = 10^{-1}\ cm^2/s$ for all $F_0$ values.

To obtain an overall impression of the size of a vortex ring core an *equivalent core radius*, $R_c$, is defined as follows:

$$R_c = \sqrt{\Gamma/(\pi\Omega_m)} \quad , \tag{19}$$

where $\Gamma$ is the total flow circulation defined by equation (12) and $\Omega_m$ is the maximum vertical vorticity in the $x - z$ plane in which the circulation is determined. We will refer to this subsequently as the core radius. This can be thought of as the radius of a circle which contains the total flow circulation but with a uniform vorticity $\Omega_m$. In Figure (8A) the evolution of the core radius is shown for the lowest viscosity case and for the entire duration of the simulation. In general, the core radius is seen to increase in time. However, near the end of region T1 ($t{\sim}0.2s$) oscillations in the core radius are observed, particularly for the highest force amplitudes. These oscillations appear to correspond to the coalescence of the vorticity and the instability of the shear layer (see Figure 4) in the early stages of the formation of each vortex ring. In Figure (8B) the core radius normalized by the radius of the body force, $R^* = R_c/R$, versus $t^*$ is shown for the region T1 for three different viscosities. Of particular note is the time at which the core radius is approximately equal to the radius of the body force, $R^*{\sim}1$. We can see from these results that as the fluid viscosity increases, the time at which this occurs also increases. This observation will be of importance in developing a model which can account for viscous diffusion.

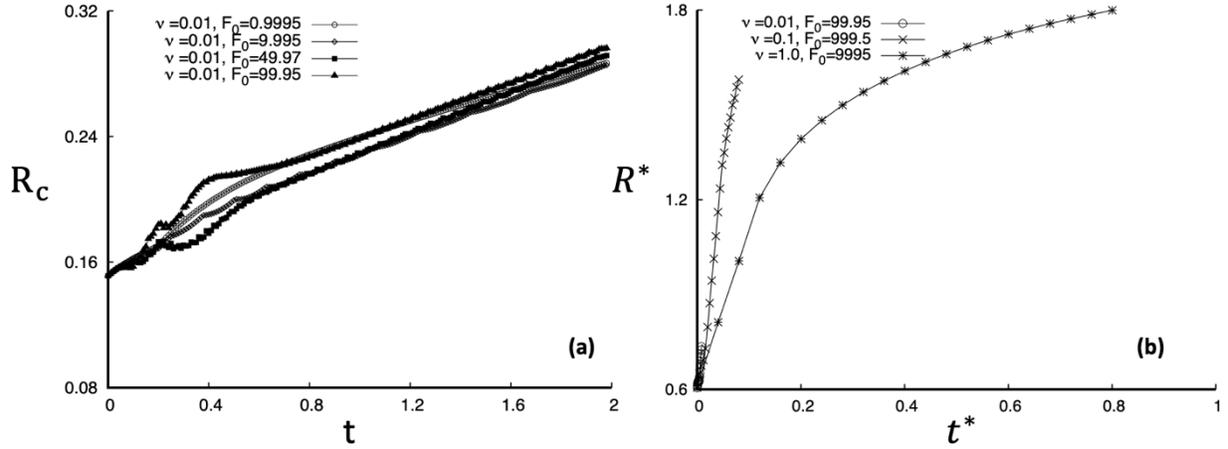

**Figure 8**: (A) Equivalent core radius per ($R_c$ in cm) versus time in seconds for the cases shown in the graph. T1 and T2 are shown. The force turns off at $t = 0.2s$. (B) Non-dimensional equivalent core radius ($R^* = \frac{R_c}{R}$) versus $t^*$ for all cases shown in the graph for T1.

Results for the volume average kinetic energy, $K$, are shown in Figure 9. In Figure 9(A), which shows the entire history of the kinetic energy in T1 and T2, the rate of change of $K$ is seen to change from positive to negative at the instant the body force is turned off at $t = 0.2\ s$. This is consistent with the prediction of equation (17). Although only two cases are shown, this was found to be true in all cases. To obtain a rough idea of the nature of the velocity distribution in the flow, the evolution of a non-dimensional velocity, $K^* = \sqrt{K/F_0 l_f}$, is shown in Figure 9(B), for typical cases in T1. For these typical cases, $K^* \lesssim 8 \times 10^{-2}$, which shows that the flow kinetic energy is concentrated in a small region associated with the vortex ring since $\sqrt{K}$ gives an estimate of the average magnitude of the flow velocity in the entire domain, and $\sqrt{F_0 l_f}$ estimates the maximum flow velocity near the ring.

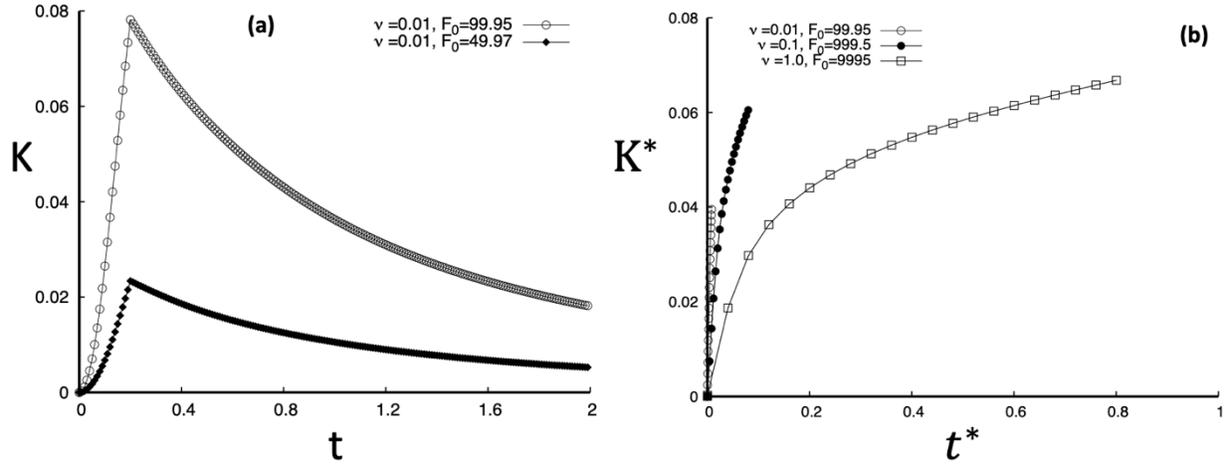

**Figure 9**: (A) Kinetic energy per unit mass (K in $cm^2/s^2$) versus time in seconds for the cases shown in the graph doe T1 and T2. (B) $K^* = \sqrt{\dfrac{K}{F_0 l_f}}$ versus $t^*$ for three cases in T1.

# 6. Models for the circulation of laminar vortex rings generated by an impulsive body force and comparison with numerical simulations

## 6.1 Model problem for the vorticity

An objective of this work is to develop a model which can be used to estimate the circulation of laminar vortex rings generated by an impulsive force. The simplest model would be one which takes into account viscous diffusion as described by equation (14) and neglects nonlinearity. The extent to which this is approximation is accurate will be determined by comparing the result of such a model with the DNS. To this end, we consider the following model equation for a scalar, $\varphi$, which represents the vertical vorticity in our problem:

$$\frac{\partial \varphi}{\partial t} = \nu \nabla^2 \varphi + S, \qquad (20)$$

where the source $S$ represents the rate at which vorticity is generated by body forces. Further simplification is obtained by applying boundary conditions $\lim_{x,y,z \to \pm\infty} \varphi = 0$, and the initial condition $\varphi(t = 0) = 0$. The exact solution to equation (20) with the prescribed boundary and initial conditions can be obtained by using Fourier transform methods [23] as follows:

$$\varphi = \int_0^t \iiint_{-\infty}^{+\infty} (\pi\sigma)^{-3/2}\, S(\varepsilon,\eta,\gamma) e^{\frac{-(x-\varepsilon)^2}{\sigma}} e^{\frac{-(y-\eta)^2}{\sigma}} e^{\frac{-(z-\gamma)^2}{\sigma}}\, d\varepsilon d\eta d\gamma d\bar{t} \quad (21)$$

where $\sigma = 4\nu(t - \bar{t})$, and $S(\varepsilon,\eta,\gamma)$ does not depend on time during T1, consistent with the source used in the numerical simulations. The solution given by equation (21) should therefore be valid only for the time period in which the force acts.

In Section 2 the actual force field was described. Based on that description, a reasonable approximation for $S$ for which exact integrals can be obtained is as follows:

$$S = X(\varepsilon)Y(\eta)Z(\gamma) \quad , \quad (22)$$

where $X(\varepsilon)$, and $Y(\eta)$ are zero except in the region $-l_f/2 \leq \varepsilon \leq -l_f/2$, $-R \leq \eta \leq R$ in which $X(\varepsilon) = Y(\eta) = 1$. The function $Z(\gamma)$ is given by:

$$Z(\gamma) = F_0[\delta(\gamma - R) - \delta(\gamma + R)] \,, \quad (23)$$

where the delta functions arise from the differentiation of a discontinuous force field model. This would arise, for example, in the limit of unbounded tapering parameter, $\beta \to \infty$. We refer to this model as a *slab* model, in the sense that the source of the vertical component vorticity exists only in a rectangular region, or *slab*. This is an approximation for the actual source of vorticity (see Figure 3), which is somewhat more complex than suggested by the slab model since the force itself exists in a cylindrical region. Nevertheless, its use will be shown to predict the vortex ring circulation in reasonable agreement with results obtained from the actual source.

### 6.2. Solution for the vorticity and the circulation obtained from the slab model

Here we are interested in the vertical vorticity on the center plane, $\varphi_0 = \varphi(y = 0)$, since this will determine the circulation according to equation (12). This is obtained by substituting equations (22,23) for the source into equation (21) and performing the appropriate integrals. This leads to the following result for $\varphi_0$:

$$\varphi_0 = F_0 \int_0^t (\pi\sigma)^{-3/2} f(x) h(\sigma) g(z) \, d\bar{t}, \tag{24}$$

where

$$f(x) = \frac{\sqrt{\pi\sigma}}{2}\left[\text{erf}\left(\frac{x + l_f/2}{\sqrt{\sigma}}\right) - \text{erf}\left(\frac{x - l_f/2}{\sqrt{\sigma}}\right)\right], \tag{25}$$

$$g(z) = e^{\frac{-(z-R)^2}{\sigma}} - e^{\frac{-(z+R)^2}{\sigma}}, \tag{26}$$

and

$$h(\sigma) = \sqrt{\pi\sigma}\,\text{erf}\left(\frac{R}{\sqrt{\sigma}}\right). \tag{27}$$

The circulation, $\Gamma$, is obtained as follows:

$$\Gamma = \int_0^\infty \int_{-\infty}^\infty \varphi_0 \, dx \, dz \tag{28}$$

Substitution of equation (24) into equation (28) and carrying out the integrals gives:

$$\Gamma = F_0 l_f \int_0^t \left(\text{erf}\left(\frac{R}{\sqrt{\sigma}}\right)\right)^2 d\bar{t} \tag{29}$$

The integral in equation (29) cannot be obtained in terms of known functions. Nevertheless, some understanding of the nature of the circulation can be had by considering the argument of the error function, $R/\sqrt{\sigma}$, which can be interpreted as the ratio of the force radius to a viscous layer thickness, $\delta_v \sim \sqrt{4\nu t}$. It follows that owing to the asymptotic behavior of the error function, $\lim_{\zeta \to \infty} \text{erf}(\zeta) = 1$, that:

$$\lim_{R \gg \delta_v} \Gamma = F_0 l_f t \tag{30}$$

This shows that when the viscous layer has not grown to a sufficient extent, the circulation grows linearly in time, in agreement with the inviscid result given by equation (15). In this case, the magnitude of the circulation depends only on the product of the force amplitude and the longitudinal ($x$) extent of the force. In the next section, we compare the results of the direct simulations to the inviscid result and the slab model. Comparisons will also be made with spectrally accurate solutions of the diffusion equation (20) using the exact source of vorticity, $\frac{\partial f_x}{\partial z}$, where $f_x$ is defined by equations (3-5).

### 6.3 Comparison of the inviscid, slab, and diffusion models with direct numerical simulations

In Figure (10) the results of the direct numerical simulations (DNS) for the temporal evolution in T1 of the vortex ring Reynolds number, $Re_\Gamma$, are compared with the inviscid, slab, and diffusion models for three viscosities. The results for the slab model were obtained by numerically integrating equation (28). The results for the *diffusion model* are obtained from a spectrally accurate solution of the *linear* diffusion equation [equation (20)] for a passive scalar, $\varphi$, with the same source, $S$, used to generate the vortex ring. In this model the no-flux condition, $\partial \varphi / \partial x_2 = 0$, is applied to top and bottom walls of the computational domain, which is the same domain used to simulate vortex rings. Note that this condition is consistent with the boundary condition applied to the vertical vorticity for ring simulations, but for the diffusion model, there was no flow.

It should be emphasized that the direct numerical simulations include the full nonlinear equations of motion, which naturally include vortex stretching, tilting, and advection of vorticity as embodied in equation (10). Furthermore, it is important to reiterate that the inviscid model should be accurate only when $t \ll R^2/4\nu$ or in non-dimensional form as $t^* \ll R^2/4l_f^2$. However, we can obtain a more accurate estimate of the time at which viscous diffusion begins to affect the circulation by using the slab model (equation (28)) as a guide. This can be done by setting $erf^2(q) = 0.95$ from which we obtain $t^*_{inv} = R^2/4q^2 l_f^2 = 2.5 \times 10^{-2}$, which we call the inviscid time.

For the lowest viscosity case shown in Figure (10a), it is clear that all three models agree well with the DNS. In fact, all models predict essentially inviscid behavior since the maximum time in this case is much less than $t_{inv}^*$. For the intermediate viscosity shown in Figure 10(b), the diffusion model shows excellent agreement with the DNS throughout the entire time interval. The slab model clearly exhibits the effects of viscous diffusion, and shows reasonable agreement with the DNS, while the inviscid model shows the greatest deviation from the numerical simulations. For the highest viscosity shown in Figure 10(c), the diffusion and slab models again show good agreement with the DNS. In summary, we note that in all cases when $t^* < t_{inv}^*$, the inviscid model agrees well with the DNS. Furthermore, we find that the effects of viscous diffusion become evident when $t^* > t_{inv}^*$ for slab and diffusion models, as well as for the direct numerical simulations.

The excellent agreement between the diffusion model and the DNS indicates that viscous diffusion alone can account for the evolution of laminar vortex rings generated by an impulsive body force for the range of Reynolds numbers, $Re_f$, investigated in this work. Furthermore, the slab model, though not as accurate as the diffusion model, can also be used to predict the ring circulation with a reasonable degree of accuracy. The fact that a pure diffusion model alone can accurately predict the evolution of the circulation requires further explanation. Consider, for example, a purely two-dimensional vortex such as the Lamb-Oseen vortex [24]. In this case, since the flow is two-dimensional and there is no source of vorticity, the circulation remains constant in time, although vorticity does diffuse in the plane of the vortex. However, in the simulations of laminar vortex rings performed in this work, lateral and vertical diffusion cannot be neglected because of the three-dimensional geometry of the vortex ring. In fact, the slab model predicts that for $t^* > t_{inv}^*$, viscous diffusion is important in predicting the evolution of the circulation.

Moreover, these results imply that as far as the circulation is concerned, nonlinearity associated with vortex stretching and tilting are apparently negligible since they are not accounted for in the diffusion model.

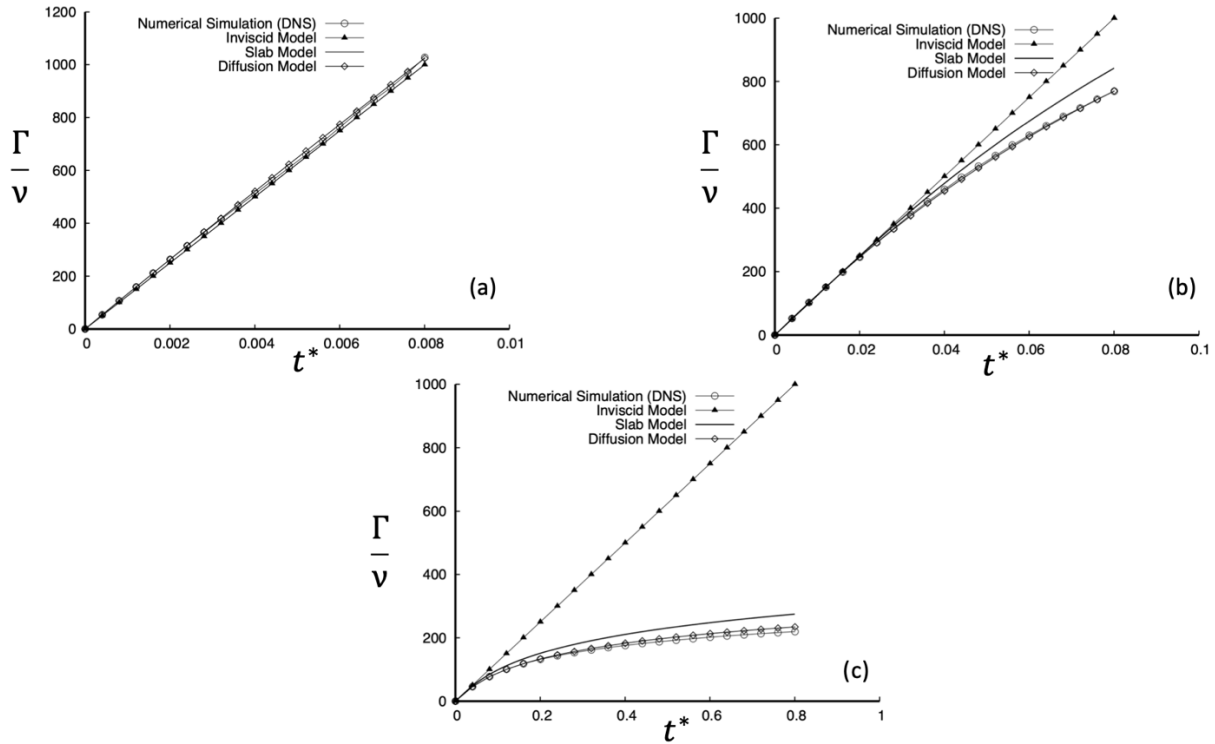

**Figure 10**: The vortex ring Reynolds number, $Re_\Gamma = \Gamma/\nu$, versus $t^*$ for the direct numerical simulation, inviscid model, slab model, and diffusion model in T1. (A) $\nu = 10^{-2}\ cm^2/s$ and $F_0 = 99.95\ cm/s^2$. (B) $\nu = 10^{-1}\ cm^2/s$ and $F_0 = 999.95\ cm/s^2$. (C) $\nu = 1.0\ cm^2/s$ and $F_0 = 9999.5\ cm/s^2$. Note that when $t^* < t^*_{inv}$ where $t^*_{inv} = 2.5 \times 10^{-2}$, the effects of viscous diffusion are negligible for slab and diffusion models, as well as the direct numerical simulations.

### 6.4 Schematic representation of the lateral diffusion of vorticity

We have shown that models based on the diffusion of vorticity can be used to accurately predict the evolution of the circulation for short as well as long time scales. These models account for both vertical as well as lateral diffusion. To obtain a qualitative understanding of this, it is useful to consider the effects of lateral diffusion alone, from which we can gain insight into the physical significance of the ratio $R/\delta_\nu$, the ratio of the force radius to the viscous diffusion length scale. To this end, we show in Figure 11 a schematic illustrating the temporal evolution of the lateral ($z$) diffusion of vertical vorticity in the time period T1. In Figure 11(A) regions of positive (red) and negative (blue) vertical vorticity are shown as they appear immediately after the initiation of the body force. At this time, the viscous diffusion length scale is negligibly small compared to $R$. An image virtually identical to this appears in the simulations as shown in Figure 4(A). At an intermediate time shown in Figure 11(B), the diffusion length scale becomes evident as vorticity is seen to diffuse from both positive (red) and negative (blue) regions. Here we note that positive vorticity is seen to diffuse in both positive and negative z directions, and this is also true the region of negative vorticity. In Figure 11(C) the vorticity is seen to have diffused to a scale $\delta_v \sim R$ at a time $t \sim R^2/4\nu$. At this time, the upper half plane, which is the region over which circulation is obtained, is seen to lose positive vorticity and gain negative vorticity. This net negative flux of vorticity across the symmetry plane results in a reduction of the rate of increase of circulation as predicted by the slab model. It should also be kept in mind that vertical diffusion of vorticity, not depicted here, also contributes to the decreased rate of increase of circulation.

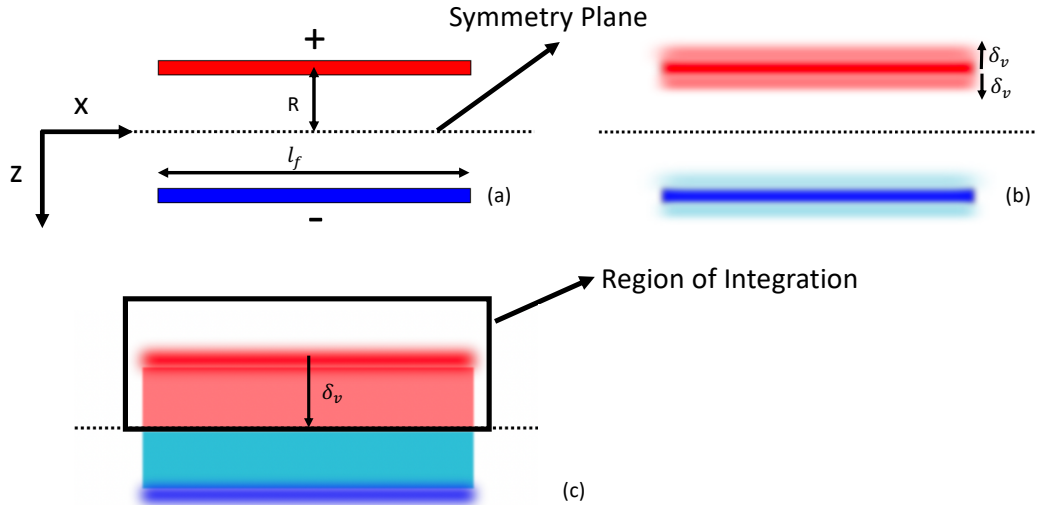

**Figure 11**: Schematic representation of the lateral diffusion of vertical vorticity in T1. (A) Representation of the positive (red) and negative (blue) vertical vorticity layers immediately after the force acts on the fluid. This corresponds to the vorticity field shown in Figure 4(a). (B) The vorticity is shown at a later time where it is seen to diffuse in the positive and negative z-direction from the red region. The viscous layer thickness is $\delta_v$. (C) The vorticity is shown at a time $t \sim R^2/4\nu$. Here the vorticity is shown diffusing from the red layer in the positive z-direction. The viscous layer thickness has reached the symmetry plane such that its thickness is approximately equal to the force radius, $\delta_v \sim R$. The circulation is determined by integrating the vertical vorticity over the region of integration shown.

## 7. Summary and Discussion

Vortex rings can be generated by a variety of experimental methods, the most recent of which involves the use of a piston-driven system [3]. In contrast to these experimental methods, we explore the possibility of generating rings using impulsive body forces. Our main objective is to develop a physical understanding of the ring generation process and thereby to develop models that can be used to predict the strength of laminar vortex rings generated in this way. We find, through direct numerical simulations, that laminar vortex rings can be easily generated by a force whose amplitude, duration, and geometry are appropriately chosen. In particular, a force which exists in a cylindrical region, similar to the piston geometry used in experiments, produces vortex rings of the desired shape and strength. Furthermore, by comparing the results of simulations with analytical models, we show that it is possible to accurately predict the strength of laminar rings by considering only the diffusion of vorticity, in spite of the importance of nonlinearity in the development of the ring.

The main result of this work can be summarized by $Re_\Gamma = Re_f^2 T(t^*)$, which states that for laminar vortex rings generated by an impulsive body force, the vortex ring Reynolds number is the product of the square of the force-based Reynolds number and a function of a viscosity-based non-dimensional time. Simulations show that this result is accurate over a range of two orders of magnitude in $Re_f$. Moreover, this also agrees with both an analytical model (slab model), based on an approximation of the source of vorticity, and spectrally accurate solutions of the diffusion equation using the exact source. Models and simulations also show that when $t^* \ll R^2/4l_f^2$, which states that the force acts for a time short enough for the diffusion of vorticity to be negligible, that $\Gamma/\nu = Re_f^2 t^*$. We also emphasize the idea that since a vortex ring is manifestly a three-dimensional vortical structure, that both lateral and vertical diffusion of vorticity must be taken into account in order to accurately predict the ring circulation when $t^* > t_{inv}^*$ where $t_{inv}^* = R^2/4q^2 l_f^2$. This differs from the case of two-dimensional vortices.

It is also important to reiterate that the use of impulsive body forces to generate coherent vortices such as rings, may have advantages compared with other methods. Flows generated in this way in a fluid initially at rest will automatically satisfy boundary conditions and will be divergence-free. Furthermore, although this work has focused on laminar rings, we expect that the same methods can be used to generate turbulent rings [11]. The simplest way to achieve this would be to generate a ring within the inviscid regime such that the force acts for a time $\tau \sim R^2/4\nu$. This will produce a freely propagating ring with a large enough Reynolds number $Re_\Gamma \sim F_0 l_f \tau/\nu$ to produce a turbulent ring by choosing an appropriate force amplitude. Finally, this work leaves open the question of whether impulsive body forces can be generated experimentally, for example, by using Lorenz forces. This would allow further verification of the results presented in this work.

## Acknowledgements:

Support was provided by the National Science Foundation under grants 1904953 and 1905288. Simulations were performed on ARGO, a research computing cluster provided by the Office of Research Computing at George Mason University. Additional support for R.K. was provided by the National Science Foundation Graduate Research Fellowship under grant DGE 2137420.

## Data Availability Statement:

The data that support the findings of this study are available from the corresponding author upon reasonable request.

# Appendix A: Initial Time-Rate-of-Change of the Vorticity

Here we give further details regarding the derivation of equation (8), which relates the initial time-rate-of-change of vorticity to spatial gradients of the body force field. Refer to the text or Table1 in Appendix B for the definitions of all symbols not explicitly defined in this Appendix. Here it is assumed that the fluid is initially at rest ($\boldsymbol{V} = 0$ at time $t = 0$ ) and therefore the vorticity $\boldsymbol{\Omega} = \boldsymbol{\nabla} \times \boldsymbol{V}$ is also zero at $t = 0$.

For convenience we rewrite the evolution equation for the vorticity:

$$\frac{D\boldsymbol{\Omega}}{Dt} = (\boldsymbol{\Omega} \cdot \boldsymbol{\nabla})\boldsymbol{V} + \nu\nabla^2\boldsymbol{\Omega} + \boldsymbol{\nabla} \times \mathbf{f} , \qquad (A1)$$

The first term on the left-hand side of equation (A1), which represents the material derivative of the vorticity, can be written, using index notation in Cartesian coordinates as follows:

$$\frac{D\Omega_m}{Dt} = \frac{\partial \Omega_m}{\partial t} + V_i \partial_i \Omega_m , \qquad (A2)$$

where $\partial_i = \partial/\partial x_i$ , and repeated indices imply summation. Evaluation of equation (A2) at $t = 0$ and using assumption (1) above gives:

$$\left.\frac{D\,\Omega_m}{Dt}\right|_{t=0} = \left.\frac{\partial\,\Omega_m}{\partial t}\right|_{t=0} . \qquad (A3)$$

The second term on the right-hand side of equation (A2) vanishes since the fluid is initially at rest, and for the same reason the first and second terms on the right-hand side of equation (A1) also vanish. As a result, equation (A1) reduces to:

$$\frac{\partial \Omega_m}{\partial t} = \frac{\partial f_j}{\partial x_i}\epsilon_{mij}, \qquad (A4)$$

at $t = 0$. However, since the force field we consider has only one non-zero component such that $\boldsymbol{f} = \boldsymbol{f}(\boldsymbol{x}, t) = (f_1, 0, 0)$, equation (A4) becomes:

$$\frac{\partial \Omega_2}{\partial t} = \frac{\partial f_1}{\partial x_3}, \text{ and } \frac{\partial \Omega_3}{\partial t} = -\frac{\partial f_1}{\partial x_2}, \qquad (A5)$$

at $t = 0$. This corresponds exactly to equation (8) since subscripts (1, 2, 3) correspond to the coordinates $x$, $y$, and $z$ respectively. We note that equation (A5) states that although the initial vorticity is zero, the initial time-rate-of-change of vorticity is not zero due to action of the body force.

## **Appendix B: Evolution Equation for the Circulation**

Here we give further details regarding the derivation of equation (12), the evolution equation for the circulation around a laminar vortex ring. The vertical vorticity, $\Omega_y$, is governed by:

$$\frac{\partial \Omega_y}{\partial t} = \frac{\partial H_x}{\partial z} - \frac{\partial H_z}{\partial x} + \nu\nabla^2\Omega_y + \frac{\partial f_x}{\partial z}, \qquad (B1)$$

where $H_x = v\Omega_z - w\Omega_y$, and $H_z = u\Omega_y - v\Omega_x$. To obtain the desired evolution equation for the circulation defined by:

$$\Gamma = \overline{\Omega_y(x, 0, z)} \equiv \int_0^{l_z/2}\int_0^{l_x} \Omega_y(x, 0, z, t)\,dx\,dz, \qquad (B2)$$

where $\Omega_y(x, 0, z)$ is the vertical component of vorticity on the center plane, we apply the overbar operation to equation (B1) which gives:

$$\frac{\partial \Gamma}{\partial t} = \overline{\frac{\partial H_x}{\partial z}}\bigg|_{y=0} - \overline{\frac{\partial H_z}{\partial x}}\bigg|_{y=0} + \nu\left[\overline{\frac{\partial^2\Omega_y}{\partial x^2}}\bigg|_{y=0} + \overline{\frac{\partial^2\Omega_y}{\partial y^2}}\bigg|_{y=0} + \overline{\frac{\partial^2\Omega_y}{\partial z^2}}\bigg|_{y=0}\right] + \overline{\frac{\partial f_x(x, 0, z, t)}{\partial z}}. \qquad (B3)$$

In our simulations, all field variables are expanded in Fourier modes in the horizontal $(x - z)$ and therefore satisfy periodic boundary conditions in these two directions. This immediately gives:

$$\left.\frac{\overline{\partial H_z}}{\partial x}\right|_{y=0} = \left.\frac{\overline{\partial^2 \Omega_y}}{\partial x^2}\right|_{y=0} = 0. \tag{B4}$$

Furthermore, the first term on the right-hand side of equation (B3) can be written:

$$\left.\frac{\overline{\partial H_x}}{\partial z}\right|_{y=0} = \int_0^{l_x}[v\left(x,0,\tfrac{l_z}{2}\right)\Omega_z\left(x,0,\tfrac{l_z}{2}\right) - w\left(x,0,\tfrac{l_z}{2}\right)\Omega_y\left(x,0,\tfrac{l_z}{2}\right) - v(x,0,0)\Omega_z(x,0,0) + w(x,0,0)\Omega_y(x,0,0)]dx \quad .$$

(B5)

However, since the symmetry of the body force imposes symmetry on the laminar vortex rings, the vertical and spanwise components of velocity, $v\left(x,0,\tfrac{l_z}{2}\right), w\left(x,0,\tfrac{l_z}{2}\right), v(x,0,0), w(x,0,0)$ are all zero, and it follows that that $\left.\frac{\overline{\partial H_x}}{\partial z}\right|_{y=0} = 0$. Applying the overbar operation to the remaining viscous terms gives $D = \nu \overline{(\Omega_y)}_{yy}\big|_{y=0}$, which represents the vertical ($y$) diffusion of vertical vorticity, and $L = \nu \int_0^{l_x} \left[\left.\frac{\partial \Omega_y}{\partial z}\right|_{z=L_z/2} - \left.\frac{\partial \Omega_y}{\partial z}\right|_{z=0}\right]\bigg|_{y=0} dx$ which represents the net lateral flux of vertical vorticity. This results in an equation for the evolution of the circulation given by:

$$\frac{\partial \Gamma}{\partial t} = D + L + \frac{\overline{\partial f_x(x,0,z,t)}}{\partial z} \quad . \tag{B5}$$

Equation (B6) corresponds to equation (12). It is important to note that terms $D$ and $L$ in equation (B6) do not contain the circulation $\Gamma$, but instead contain the vertical vorticity and its derivatives. Therefore, in spite of the linear appearance of equation (B6), nonlinearity can enter the evolution equation for the circulation through the nonlinear evolution equation for the vertical vorticity given by equation (B1).

The third term on the right-hand side of equation (B6) is the source of the circulation. To get an idea of the nature of this term, we find that $\int_0^{l_z/2} \int_0^{l_x} \frac{\partial f_x}{\partial z} dx dz = 0.4994$ using a $128 \times 65 \times 128$ grid, a tapering parameter $\beta = 15\ cm^{-1}$, $l_f = 2R = 0.5\ cm$, and $F_0 = 1$. This is virtually identical to the result we would get had the force been described by: $f_x = 1$ for $-l_f/2 \leq x \leq$ $-l_f/2$ and $-R \leq z \leq R$ , and 0 otherwise. This implies that the force field is accurately modeled as one which has sharp spatial cutoffs (e.g., it is either 0 or 1). The force, $f_x$ , and its spanwise derivative, $\frac{\partial f_x}{\partial z}$ , are shown in Figures 1 and 3.

# Appendix C: Table of Symbols

**Table1:** This table lists all relevant symbols used in this work. For each symbol, a definition is given followed by a designation of the symbol as dimensional (e.g., centimeters, seconds) or dimensionless.

| Symbol | Definition | Units |
|---|---|---|
| $x, y, z$ | Coordinates: the vortex ring travels in the x-direction, z is perpendicular to this direction, and y is the vertical direction (see Fig. 1) | Dimensional |
| $x_1, x_2, x_3$ | Scripted coordinates corresponding to $x, y, z$ | Dimensional |
| $\mathbf{x}$ | Position vector | Dimensional |
| $l_x, l_y, l_z$ | Lengths of the computational domain in the $x$ , $y$ , and $z$ directions. $l_y$ is the half-height of the computational domain in the y-direction | Dimensional |
| $L_x, L_y, L_z$ | Computational domain lengths made dimensionless using $l_y$. Note that this definition leads to $L_y = 2$. | Dimensionless |
| $t$ | Time | Dimensional |
| $\mathbf{V}, V_i$ | Velocity and components of the velocity | Dimensional |
| $\mathbf{\Omega}, \Omega_i$ | Vorticity and components of the vorticity | Dimensional |
| $u, v, w$ | Components of the velocity in the $x, y$, and $z$ directions | Dimensional |
| $u_1, u_2, u_3$ | Scripted components of the velocity | Dimensional |
| $\Omega_x, \Omega_y, \Omega_z$ | Components of the vorticity in the $x, y$, and $z$ directions | Dimensional |
| $\Gamma$ | Circulation | Dimensional |
| $p$ | Pressure | Dimensional |
| $\rho$ | Density | Dimensional |
| $f$ | Body force | Dimensional |
| $f_x, f_y, f_z$ | Components of the body force in the $x, y$, and $z$ directions | Dimensional |
| $f_1, f_2, f_3$ | Scripted components of the body force | Dimensional |
| $F_0$ | Magnitude of the body force | Dimensional |
| $\tau$ | Temporal duration of the body force | Dimensional |
| $x_0, y_0, z_0$ | Coordinates of the center of the domain and also coordinates of the of the center of the body force | Dimensional |
| $A(t)$ | Function used to define the temporal evolution of the body force | Dimensional |
| $r$ | Radial coordinate used to define the body force | Dimensional |
| $g_r, g_x$ | Spatial functions used to describe the body force | Dimensional |
| $R$ | Body force radius | Dimensional |
| $l_f$ | Body force length | Dimensional |
| $\beta$ | Tapering parameter | Dimensional |
| $H_x, H_z$ | Nonlinear terms in the evolution equation for the vertical vorticity | Dimensional |
| $D, L$ | Viscous diffusion terms in the evolution equation for the circulation | Dimensionless |
| $\varphi$ | Scalar field | Dimensional |
| $\varphi_0$ | Scalar field on the center plane in the slab model | Dimensional |
| $S$ | Source of the scalar field | Dimensional |
| $\sigma$ | Viscous decay function in diffusion model | Dimensional |
| $\nu$ | Kinematic viscosity | Dimensional |
| $Re_f$ | Force-based Reynolds number | Non-dimensional |
| $t^*$ | Viscous-based time scale | Non-dimensional |
| $t^*_{inv}$ | Inviscid time scale | Non-dimensional |
| $Re_\Gamma$ | Vortex ring Reynolds number | Non-dimensional |
| $t_{Ro}$ | Roshko number | Non-dimensional |
| $k$ | Kinetic energy per unit mass | Dimensional |
| $K$ | Volume average kinetic energy per unit mass | Dimensional |
| $K^*$ | Non-dimensional velocity scale | Non-dimensional |
| $R^*$ | Non-dimensional equivalent vortex core radius | Non-dimensional |
| $R_c$ | Equivalent core radius | Dimensional |
| $\epsilon$ | Volume average dissipation of kinetic energy | Dimensional |
| $\Delta x, \Delta y, \Delta z$ | Computational grid resolution | Dimensional |
| $\Omega_m$ | Maximum vertical vorticity in the horizontal plane | Dimensional |
| $\delta_v$ | Viscous layer thickness | Dimensional |
| $\Gamma^*$ | Dimensionless circulation | Non-dimensional |
| $\varepsilon, \eta, \gamma$ | Variables of integration | Dimensional |
| $f, g, h$ | Functions used in slab model | Dimensional |

# Appendix D: Simulation Parameters

**Table2:** This table lists the parameters used in all simulations. For each run, the force amplitude, $F_0$, kinematic viscosity, $\nu$, and force-based Reynolds number, $Re_f = \frac{v^* l_f}{\nu}$ are listed. Here $v^* = \sqrt{F_0 l_f}$, and $l_f$ is the length of the cylindrical region occupied by the body force.

| Run Number | $F_0 (cm/s^2)$ | $\nu (cm^2/s)$ | $Re_f$ |
|---|---|---|---|
| Run 1 | 0.9995 | 0.01 | 35.346 |
| Run 2 | 9.995 | 0.01 | 111.775 |
| Run 3 | 49.975 | 0.01 | 249.937 |
| Run 4 | 99.95 | 0.01 | 353.465 |
| Run 5 | 9.995 | 0.1 | 11.1775 |
| Run 6 | 99.95 | 0.1 | 35.346 |
| Run 7 | 499.75 | 0.1 | 79.037 |
| Run 8 | 999.5 | 0.1 | 111.775 |
| Run 9 | 99.95 | 1.0 | 3.5346 |
| Run 10 | 999.5 | 1.0 | 11.1775 |
| Run 11 | 4997.5 | 1.0 | 24.9937 |
| Run 12 | 9995 | 1.0 | 35.3465 |

# Appendix E: Evolution Equation for the Kinetic Energy

The equation for the evolution of the volume averaged kinetic energy given in equation (17) is derived below. Taking the dot product of the velocity **V** with equation (1) and using incompressibility, $\nabla \cdot \mathbf{V} = \mathbf{0}$, gives:

$$\frac{\partial k}{\partial t} = -\partial_j(ku_j) + \rho^{-1}\partial_j(pu_j) + \nu\partial_j\partial_j k - \nu\partial_j u_i \partial_j u_i + f_j u_j \quad , \tag{E1}$$

where $k = \frac{u_i u_i}{2}$. Averaging equation (E1) over the entire flow volume, V, gives:

$$\frac{\partial K}{\partial t} = -<\partial_j(ku_j)> + \rho^{-1}<\partial_j(pu_j)> + \nu<\partial_j\partial_j k> - \epsilon + <f_j u_j>, \tag{E2}$$

where the volume-average kinetic energy per unit mass is $K = <k>$ and the dissipation is $\epsilon = \nu <\partial_j u_i \partial_j u_i>$. The application of periodic boundary conditions in the horizontal $(x - z)$ plane and shear free conditions on the upper and lower domain boundaries leads to the elimination of the first three terms on the right-hand side of equation (E2). The final equation for the volume averaged kinetic energy, $K$, is therefore:

$$\frac{\partial K}{\partial t} = -\epsilon + <f_j u_j>, \tag{E3}$$

which is equation (17) in the main text.